# IOT NETWORK BEHAVIORAL FINGERPRINT INFERENCE WITH LIMITED NETWORK TRACES FOR CYBER INVESTIGATION : A META LEARNING APPROACH

JONATHAN PAN

Home Team Science and Technology Agency, Singapore
E-MAIL: Jonathan_Pan@htx.gov.sg / JonathanPan@ntu.edu.sg

**Abstract:**

The development and adoption of Internet of Things (IoT) devices will grow significantly in the coming years to enable Industry 4.0. Many forms of IoT devices will be developed and used across industry verticals. However, the euphoria of this technology adoption is shadowed by the solemn presence of cyber threats that will follow its growth trajectory. Cyber threats would either embed their malicious code or attack vulnerabilities in IoT that could induce significant consequences in cyber and physical realms. In order to manage such destructive effects, incident responders and cyber investigators require the capabilities to find these rogue IoT and contain them quickly. Such online devices may only leave network activity traces. A collection of relevant traces could be used to infer the IoT's network behavioral fingerprints and in turn could facilitate investigative find of these IoT. However, the challenge is how to infer these fingerprints when there is limited network activity traces. This research proposes the novel model construct that learns to infer the network behavioral fingerprint of specific IoT based on limited network activity traces using a One-Class Time Series Meta-learner called DeepNetPrint. Our research also demonstrates the application of DeepNetPrint to identify IoT devices that performs comparatively well against leading supervised learning models. Our solution would enable cyber investigator to identify specific IoT of interest while overcoming the constraints of having only limited network activity traces of the IoT.

## 1. Introduction

The adoption of Internet of Things (IoT) will grow significantly as we enter in Industry 4.0. IoT exist in varied forms from stationary surveillance sensors to flying drones. According to World Economic Forum [1], IoT is multifaceted with a largely uncoordinated group of stakeholders that poses exciting opportunities as well as inchoate and chaotic environment. IoT are computer embedded physical devices that operates in cyber or digital space. One key feature of its cyber characteristics is their ability to communication with each other in wired or wireless communication channels to communicate exchange information and send / receive instructions. The evolution of IoT will grow along with the pervasive growth of edge computing and wireless communication technologies.

As with all technologies especially those cyber enabled or connected to the Internet, there will be Cybercriminal or Cyber Security attackers seeking opportunity to exploit the IoT's mass adoption to disrupt the security and safety of the cyber landscape. The origins of attack may start from the initial development of IoT devices with the embedding of malicious code to active adversarial attacks against vulnerable IoT like the Mirai attacks against routers and networking devices [2]. In the case of Mirai incident, there was massive disruption to transportation and health care services. Such rogue or infected IoT may exist not only in swarms or large numbers. There may be one or a few rogue IoT controlled by individual like a rogue drone. Hence against such highly probable security risks, incident responders and cyber investigators will need to have the means to identify these rogue IoT and contain them in a timely manner. However, to identify these rogue IoT, forms of datapoints that uniquely identifies these rogues or their fingerprints will need to be inferred from limitedly available traces of the rogue's activities. For IoT, that will likely be the network activity traces originating from those devices. Even if such network behavioral fingerprints may be inferred, finding these IoT is a resource intense task when these IoT exist with other non-rogue IoT and communicating devices on that same communication channel. While the daunting task of identifying the IoT may be addressed through modern techniques with the use of Machine Learning or Artificial Intelligence algorithms, however most of such algorithms requires significant amount of training data with good behavioral inductive biases to train an effective detection or identification model.

This research proposes two novelties. The first is the novel model construct that learns to infer the network behavioral fingerprint of IoT based on limited samples of network activity traces using a One-Class Time Series Meta-learner called DeepNetPrint. The second is the novel application of this DeepNetPrint to identify IoT devices of interest that performs comparatively well against leading supervised learning models. Our model enables the identification of specific IoT of interest for cyber investigation while overcoming the constraints of only having limited network activity traces of the IoT.

The next section of this paper provides background information about challenges faced in performing IoT investigation with behavioral fingerprints. This is followed by related research work in identifying IoT based on network behavioral fingerprints and in metric based meta-learner algorithms. The description of the proposed model construct is covered next, followed by the description of experiments and an analysis of the results. The paper concludes with a conclusion and discussion about future research directions.

## 2. Background Information

### 2.1. Challenges in IoT Investigation

IoT are digitally enabled forms of physical objects that could become malicious smart devices whose threat actors originate from both cyber and physical dimensions. Attacks originating from rogue IoT could have cyber and physical consequences. According to Zulkipli et al. [3], four broad categories of threat source to IoT. They are the Mischievous or Misbehave users of IoT who perform the assaults. Second is the immoral manufacturer that embeds malicious code into the devices. Third is the external adversary that exploits vulnerabilities resident in the IoT and subsequently initiate attacks on others and finally poor software development that create vulnerabilities in IoT.

When rogue IoT draws the attention of the cyber security incident responders or cybercrime investigators, the investigators' primarily task is to first find IoT devices of interest so that containment or investigation may be applied. This 'Identification' step is the first step in the Investigation Phase as defined in the Digital Forensics Investigation Process [4]. With such rogue, finding the physical location of these devices may not be easily done however there will likely be only their traces in the form of network activities available for the investigator. However, these network activity traces exist in the larger ocean of network activities from many other digitally enabled devices. Hence the problem becomes a search for the 'needle in the haystack'.

### 2.2. Cyber Investigation using Behavioral Fingerprints

Fingerprint in digital space is liken to human fingerprints that represents a cluster of datapoints that uniquely identifies an individual or object. For IoT devices or other forms of ICT equipment, such fingerprints refer to the collection of features to identify the device and its associated state [5]. The characteristics of devices' fingerprints should universally be applicable to all devices, enable unique identification of each and every device, invariant over time and facilitate collection from the devices' signals [15]. In the context of this research, the objective is to identify IoT devices of interest that carries a certain network behavioral fingerprint. Such fingerprint originates from a collection of sequence network activity traces with sufficient inductive biases to enable a model to infer its associate uniquely with the device. Such fingerprints may have unique static or sequence features that are generated from software and / or user interactions running on the computing devices.

Network behavioral fingerprints may be acquired by active or passive means that involves the use of network tools like network sniffers. However, within such network extracts or traces, they would have lots of noise in the form of network activity traces belonging to other computing nodes or devices that uses that network to communicate with each other. Hence the task of identifying whether a specific network behavioral fingerprint exists in crowded network is a complex and high laborious task if it is undertaken by manual searches.

There are solutions available in the form of signature based or machine learning / artificial intelligence algorithms to offload the complex and laborious task of identifying a network behavioral fingerprint. With signature based, their rigid signature definitions limit their ability to detect sequence based behavioral fingerprints. With machine learning or contemporary artificial intelligence algorithms, such solutions are typically constrained by the need for large number of sample fingerprints to train a model.

## 3. Related Work

### 3.1. Machine / Deep Learning in Behavioral Pattern Detection

Miettinen et al. [11] developed a solution called IoT Sentinel that is capable of identifying IP-based IoT devices based on passively observed network behavioral fingerprints originating from those devices. The classification was done using Random Forest. Aksoy and Gunes [12] further improved the identification technique with their SysID that used a variety of machine learning algorithms (Decision Table, J48 Decision Trees, OneR , and PART) with Genetic Algorithm (GA) for feature selection to automate the identification of devices from network traffic traces. However, both techniques used supervised learning approaches which may not generalize well with new devices with no prior training applied to these models.

### 3.2. Meta-Learning

Meta Learning is a subdomain of machine learning with algorithms designed to use metadata to perform automatic learning that is flexible to solve learning problems. Unlike the popular Supervised learning or Unsupervised learning algorithms that are not optimal for this task of learning from limited data samples, Meta Learning is focused on learning to learn that results in its ability to acquire knowledge versatility to learn new skills or adapt to new environment with minimal training examples. More concretely like humans, a trained meta learner model would recognize a new object with only one or a few samples. The training of Meta Learning algorithms or Meta-leaners comprises of two stages. The first is to train a classifier $f_\theta$ for a specific task. In the same time, an optimizer $S, \theta' = g_\phi(\theta, S)$ learns to update the learner's hyperparameters with support set of data from varied classes. The final step will be to update $\theta$ and $\emptyset$ to maximize the following.

$$\mathbb{E}_{\mathcal{L}\subset\mathcal{L}}\left[\mathbb{E}_{S^\mathcal{L}\subset D, B^\mathcal{L}\subset D}\left[\sum_{(x,y)\in B^\mathcal{L}} P_{g_\emptyset(\theta,S^\mathcal{L})}(y|x)\right]\right] \quad (1)$$

In this research work, we applied the metric based meta learning algorithm to address the defined problem statement of identify network behavioral fingerprints of an IoT device of interest with only one fingerprint sample. Metric based meta-learners closely resembles nearest neighbours algorithms like knn classification or k-means clustering with kernel density estimation. The predicted probability over a set of known labels $y_i$ is a weighted sum of labels of support set samples. The weight is generated by a kernel function $k_\theta$, measuring the similarity between two data samples.

$$P_\theta(y|x,S) = \sum_{(x_i,y_i)\in S} k_\theta(x, x_i) y_i \quad (2)$$

The Prototypical network [6] is one such metric based meta-learner. The algorithm computes a prototype $c$ for each class $i$ which is represented as a mean to the embeddings of the feature inputs. The Euclidean distance of the prototypes $c_i$ of the varied classes against the query is used to compute the similarity measurement.

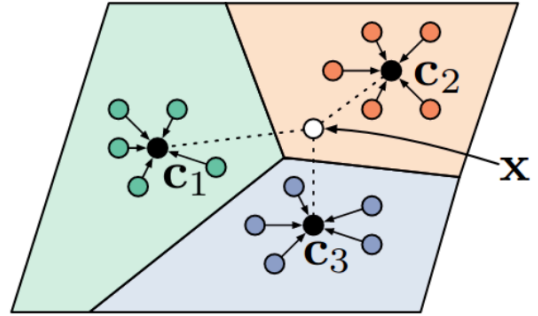

Figure 1. Extracted Diagram from Snell et. al [6]

Huang et. al [13] proposed their Deep Prototypical Networks that leveraged on the embedding space to capture discrepancies of difference time series patterns from different class to deal with problem of data scarcity.

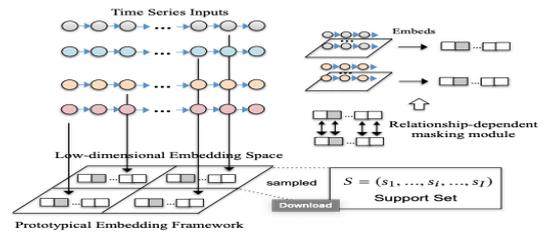

Figure 2. Extracted Diagram from Huang et. al [13]

However, ours goes further to perform similarity matches with only One Class that represents the IoT of interest instead of multiple classes.

### 4. Model

We developed a unique model construct for DeepNetPrint that is composed of two deep learning modules. The first modular construct is a ConvLSTM based Autoencoder that encodes network activity traces into encodings that represent the network activity traces embeddings. These encodings are then fed into the second modular construct which is a One Class Time Series Prototypical Network that is the meta learner based on prototype theory [14] from Cognitive Science and Prototypical Networks for few-shot classification [6].

#### 4.1. Autoencoder

The first modular construct of DeepNetPrint is an Autoencoder that uses the Convolutional Long-Short Term Memory (ConvLSTM) [7]. This ConvLSTM based Autoencoder learns the embedding of the network behavioral patterns from the input feeds extracted from the network activity traces. The embeddings used in this research is character based [8] hence the embeddings are immediately derived from raw network activity trace inputs. The ConvLSTM uses convolutional structures in both the input-to-state and state-to-state transitions of the LSTM recurrent neural network. With convolutional structures, the ConvLSTM ingests the character embeddings as spatial temporal inputs. The following are the mathematical expressions for the ConvLSTM.

$$I_t = \sigma(W_{xi} * X_t + W_{hi} * H_{t-1} + W_{ci} \circ C_{t-1} + b_i) \quad (3)$$

$$F_t = \sigma(W_{xf} * X_t + W_{hf} * H_{t-1} + W_{cf} \circ C_{t-1} + b_f) \quad (4)$$

$$C_t = f_t \circ C_{t-1} + i_i \circ \tanh(W_{xc} + X_t + W_{hc} * H_{t-1} + b_c) \quad (5)$$

$$O_t = \sigma(W_{xo} * X_t + W_{ho} * H_{t-1} + W_{co} \circ C_{t-1} + b_o) \quad (6)$$

$$H_t = o_t \circ \tanh(V_t) \quad (7)$$

The '$*$' operators that performs convolution operation while '$\circ$' is the Hadamard product for the element-wise product of matrices. The rest of network structure performs, like the LSTM network, processes the input, forget, cell, output and hidden state computations for each timestep denoted by $I, F, C, O$ and $H$ respectively with activation by $\sigma$ and convolutional filter computation with the sets of weights, $W$. Our ConvLSTM based Autoencoder can be expressed mathematically as such.

$$\alpha = x \to F \quad (8)$$

$$\beta = F \to x' \quad (9)$$

$$\alpha, \beta = \arg\min_{\alpha,\beta} \|x - (\beta \circ \alpha)x\|^2 \quad (10)$$

The training of the Autoencoder is done by minimizing the reconstruction error from the output of decoder $x'$ from the input vector $x$. However only the encoder's output $F$, that represents the feature space, is used as inputs to the next modular construct that is the One Class Time Series Prototypical Network.

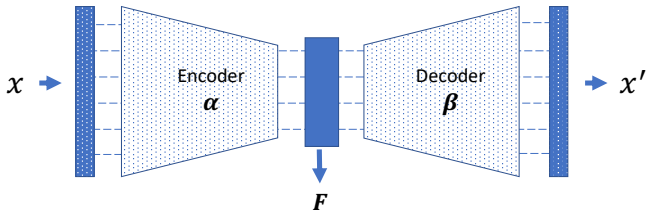

Figure 3. Autoencoder for DeepNetPrint

4.2. One Class Time Series Prototypical Networks

The second part of the modular construct of our DeepNetPrint is a One Class Time Series Prototypical Networks that performs this similarity measurement of the query point (which is a sample of the network activity traces) against the prototype of one class (or Target class) that represents the network behavioral fingerprint of the IoT device that the investigator is interested in. However, to compute the similarity comparison of the query input, there should typically be two or more classes with Prototypical Networks. We followed a similar approached used by Oza and Patel [9] by introducing a Null class represents the origin in the embedding space. The origin exists in the form of a 'silent' character embedding of the network activity traces with null characters.

Both the Target and Null classes' encodings are generated from ordered sequence of network activity trace samplings (20 sequenced samples) by the encoder of the ConvLSTM Autoencoder. These encodings are then further encoded into embedding support points by the Time Series Prototypical Network using similar approach by Huang et. al [13]. The embedded support point is computed through $f_\phi$ with learnable parameters $\phi$.

$$c_k = \frac{1}{|S_k|} \sum_{(x_i, y_i) \in S_k} f_\phi(\alpha(x_i)) \quad (11)$$

$$p_\phi(y = k|x) = softmax(-d(f_\phi(\alpha(x)), c_k))$$
$$= \frac{exp(-d(f_\phi(\alpha(x)), c_k))}{\sum_{k'} \exp(-d(f_\phi(\alpha(x)), c_{k'}))} \quad (12)$$

$$J(\phi) = -\log p_\phi(y = k|x) \quad (13)$$

The computation of the prototype $c_k$ or centroid of the class' embedded support points is the mean vector where $k$ is either Target or Null class. The measurement of the query point $x$, which is encoded by the encoder, from the network activity trace sampling is then measured as a probability distribution to the classes based on a distance function $d$ using the Euclidean distance with either of the $k$ classes.

The approach used to train of this module construct resembles the training of One-Shot Siamese Network [10] by providing labelled examples of positive and negative pairs of similarity measures in a supervised manner using the following loss function expression.

$$\delta(x^t, x^q) = \begin{cases} \max\|p_\phi(y=t|q)\|, t=q \\ \max\|p_\phi(y=n|q)\|, t\neq q \end{cases} \quad (14)$$

The objective is for the network is to maximize the match accuracies. The following diagram Figure. 4 illustrates the inputs fed into DeepNetPrint for training and inference with the same 20 sequenced network samplings from Target class that represents the Target's network behavioral fingerprint and query point that is a single network activity trace from a continuous network activity stream. The Null class is similarly fed into DeepNetPrint as a constant of 20 null samples.

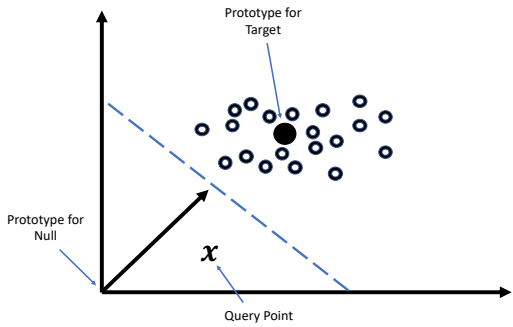

Figure 4. Prototypes and Query Point

The inputs to the combined modular constructs of both the Autoencoder and the One Class Time Series Prototypical Network or Meta-Leaner is illustrated in Figure. 5 with network inputs for the Target class, Null class and query point represented as $x_t, x_n, x_q$.

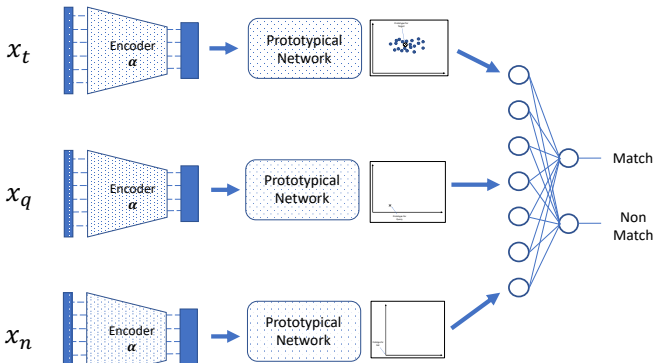

Figure 5. DeepNetPrint Framework

## 5. Methodology and Analysis

### 5.1. Dataset and Training

In our research, we used a related research's dataset of network activity packet captures (PCAP formatted files) of 23 IoT devices [11]. The devices are Aria, D-LinkCam, D-LinkDayCam, D-LinkDoorSensor, D-LinkHomeHub, D-LinkSensor, D-LinkSiren, D-LinkSwitch, D-LinkWaterSensor, EdimaxPlug1101W, EdimaxPlug2101W, EdnetGateway, HomeMaticPlug, HueBridge, HueSwitch, iKettle2, Lightify, MAXGateway, SmarterCoffee, TP-LinkPlugHS100, TP-LinkPlugHS110, WeMoLink, Withings. The packets from this dataset are network messages transmitted during the start-up sequence of the devices. From past related research work [11][12], it has been demonstrated that these packets have the needed inductive biases for our model to infer the devices' network behavioral fingerprints. The dataset was however imbalanced with variance from tens of packets to thousands from the different classes.

About half of the entire dataset, that is 12 out of 23 devices, was used for training. We trained DeepNetPrint in two phases. The first phase was to train only the Autoencoder model with the training dataset of IoT devices' network activity patterns to learn the network activity trace embeddings. The second phase involves having the encoder of the Autoencoder generate the encodings of the inputs (Target, Null and query) and feeding them into the One Class Meta-Learner with similarity pairs. The training approach used for second phase followed the training approach used by Koch et. al [10] to train a similarity classifier with positive and negative pairs. The positive pairs consist of network activity trace samplings from the same device with one of the pairs as the network behavioral fingerprint (20 sequenced samplings) representing the Target class and the other as the query point with one sample of the network activity traces. The negative pair used different network activity sampling sets from different devices for the Target class and query point. The Null class is the same for both pairs as origin to the embeddings.

The testing of the model was done with the entire dataset of 23 devices including packet data of network activity traces from the 11 of the 23 devices previously not seen by the model. The testing of the model was done to evaluate the model's ability to infer the network behavioral fingerprint of both previously trained devices and new ones to identify the devices from a single sample of network activity packet. The testing approach was organized in same manner as other research objectives in order to evaluate our model against other models namely the IoT Sentinel [11] and SysID [12].

Pre-processing was applied on the entire dataset to extract only the information source field of the PCAP into common separated values or CSV formatted files. Unlike the other referenced models that used varying number of features (up to 33) from the packets, we only extracted the 'Information Source' field containing the high-level information of the network packets that would represent 'conversational dialogues' originating from the IoT devices to train the model and infer the devices' network behavioral fingerprints identification later. This field was chosen as it contains relevant information about packets transmitted from the devices and its applicability to encrypted network activity traffic. The following extracts of datasets.

![Extract of 10 packets]

Figure 6. Extract of 10 packets from an IoT Devices from Dataset

### 5.2. Results and Analysis

Testing involved using the entire dataset that contained all 23 IoT devices' network activity samplings. The test objective was to assess DeepNetPrint's ability to infer the network behavioral fingerprint of a selected IoT device and to flag out the probable existence of the network activities originating from that Target device from a stream of network activity packets.

The procedures for the testing involved the sequential selection of each IoT device from the dataset containing network activity traces from 23 devices as Target class or device of interest. A short sample of a series of packets (20 network packets ordered sequence to represent 20 samplings) in CSV format from that device was fed to the model. The classification test was done by iterating through all CSV formatted extracts for each and every IoT device in the dataset as part of the query input packet stream $x_q$ against the Target class $x_t$ and Null class $x_n$. Both $x_t$ and $x_n$ need only be encoded once through the Autoencoder as both remain unchanged during the search for the specific fingerprint of interest. $x_q$ query stream was encoded for each and every packet. This input construct was organized in this way to closely mimic the environment that an investigator faced in assessing whether the IoT device of interest was communicating on that channel being monitored.

DeepNetPrint performed well with an identification accuracy of 81% for all 23 IoT devices and 80% for the previously unseen network activities from the 11 IoT devices.

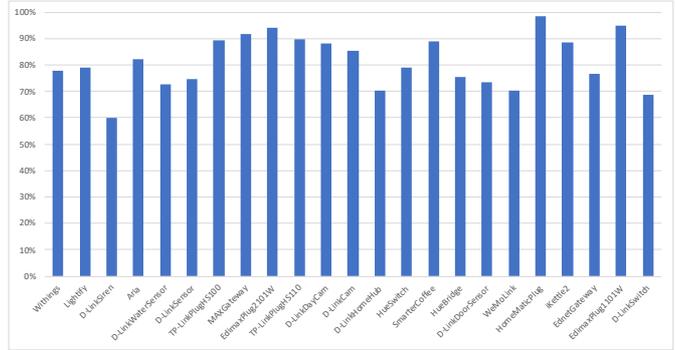

Table 1. Accuracy Performance of DeepNetPrint

We subsequently compared the accuracy performance of the DeepNetPrint with those IoT Sentinel and SysID. DeepNetFinger fared well relative IoT Sentinel's 79% and SysID's 82% for all IoT devices including the 11 devices that DeepNetPrint has not been trained before. Table 1 shows the accuracy comparison in classifying the devices between DeepNetFinger and SysID.

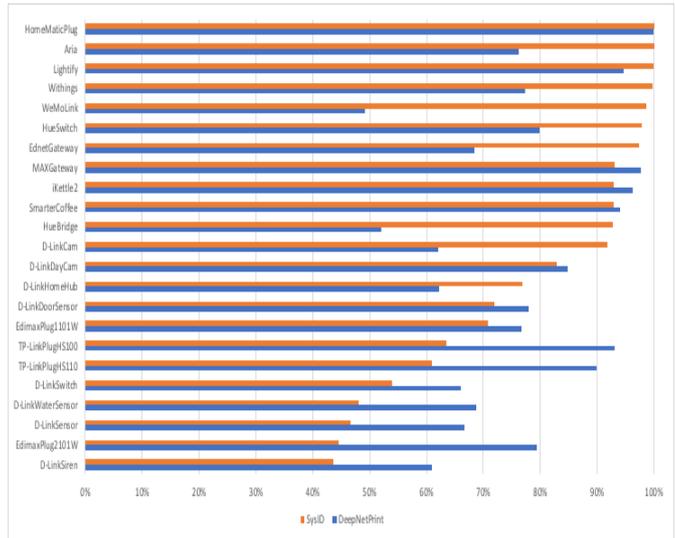

Table 1. Performance Comparison of DeepNetPrint and SysID

In some instances, DeepNetFinger performed comparatively better for devices that had similar fingerprints as they originate from the same brand or manufacturer from the dataset used.

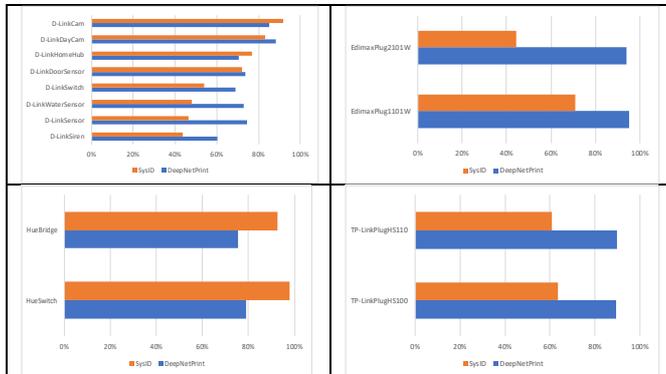

Table 2. Classification Performance of Brand Category

## 6. Conclusion and Future Directions

In this research, we developed a model called DeepNetPrint that addressed the challenge of inferring the IoT's network behavioral fingerprint with limited network activity traces. The DeepNetPrint compromises of ConvLSTM based Autoencoder and One Class Time Series Prototypical Network. We evaluated our model against other similar models that performed comparatively well in identifying IoT devices even with new devices that the model was not trained before hence generalizing well. Hence, this research work with DeepNetPrint aids cyber investigators in their challenge with identifying rogue IoT with limited network activity traces.

Future research work will focus on identifying qualifiers to infer good network behavioral fingerprint of IoT to improve identification accuracy performance. Also, to extend the application of DeepNetPrint's model construct beyond network activity traces to other forms of time series behavioral fingerprint identification.